\documentclass[prd,twocolumn,showpacs,floatfix,superscriptaddress]{revtex4}

\usepackage{graphicx}
\usepackage{amsmath}
\usepackage{amssymb}
\usepackage{bm}

\begin{document}

\title{Bulk viscosity of spin-one color superconductors with
two quark flavors}

\author{Basil A.\ Sa'd}
\email{sad@fias.uni-frankfurt.de}
\affiliation{%
Frankfurt International Graduate School for Science,
J.W. Goethe-Universit\"{a}t,
D-60438 Frankfurt am Main, Germany}%

\author{Igor A.\ Shovkovy}
\email{I-Shovkovy@wiu.edu}
\altaffiliation[on leave
       of absence from ]{%
       Bogolyubov Institute for Theoretical Physics,
       03143, Kiev, Ukraine}%
\affiliation{%
Frankfurt Institute for Advanced Studies,
J.W. Goethe-Universit\"{a}t,
D-60438 Frankfurt am Main, Germany}%
\affiliation{%
Department of Physics, Western Illinois University, Macomb, IL 61455, USA}%

\author{Dirk H.\ Rischke}
\email{drischke@th.physik.uni-frankfurt.de}
\affiliation{%
Frankfurt Institute for Advanced Studies,
J.W. Goethe-Universit\"{a}t,
D-60438 Frankfurt am Main, Germany}%

\affiliation{Institut f\"ur Theoretische Physik,
J.W.\ Goethe-Universit\"at,
D-60438 Frankfurt am Main, Germany}

\date{\today}

\begin{abstract}
We consider the contribution of the Urca-type processes to the bulk 
viscosity of several spin-one color-superconducting phases of dense 
two-flavor quark matter. In the so-called transverse phases 
which are suggested to be energetically favorable at asymptotic 
densities, the presence of ungapped quasiparticle modes prevents
that spin-one color superconductivity has a large effect on the bulk
viscosity. When all modes are gapped, as for one particular 
color-spin-locked phase, the effect on the viscosity can be 
quite large, which may have important phenomenological 
implications.
\end{abstract}

\pacs{12.38.-t, 12.38.Aw, 12.38.Mh, 26.60.+c}


\maketitle

\section{Introduction}
\label{sec0}

Dissipative processes play an important role in the evolution 
of neutron stars. These processes are governed by transport coefficients,
such as the heat and electrical conductivities, the neutrino diffusion
coefficient, as well as the shear and bulk viscosities.
The conductivities are important for stellar cooling as well as for 
the magnetic field decay. Shear viscosity dampens differential 
rotation in a star and, thus, leads to a uniform rigid-body 
rotation. Bulk viscosity, on the other hand, dampens density 
oscillations inside the star. Both differential rotation and 
oscillations could be excited in newly formed (hot) neutron 
stars, or could develop in old (cold) stars due to external 
perturbations, e.g., such as matter accreted from a companion 
star.

It is interesting to note that, in the absence of viscosity,
all rotating stars would be unstable. The reason is that such stars 
spontaneously develop instabilities as a result of the emission 
of gravitational waves \cite{Chandra1,Chandra2,Frid1,Andersson,Fried2} 
(for reviews on this topic see, e.g., Refs.~\cite{Anderssonreview,
Lind-lect}). The so-called r-mode (or rotation-dominated) 
instabilities might be the most important ones. They can develop 
at a relatively low angular velocity \cite{Lind1}, and therefore 
may be relevant for a large number of compact stars. 

The main theoretical uncertainty in predicting whether the r-mode
instabilities develop in a star lies in the poor understanding 
of the viscosity of dense baryon matter, as well as in the limited 
knowledge of the stellar composition. There is hope, however, 
that a systematic approach, based on a broad understanding of 
various properties of dense baryonic matter, can eventually 
result in a clear picture regarding the neutron star composition. 

The viscosity of nuclear and mixed phases of dense 
baryonic matter has been calculated under various conditions 
and assumptions over the last three decades \cite{FlowersItoh1,
FlowersItoh2,Sawyer,Jones1,Lindblom1,Lindblom2,Drago1,Haensel1,Haensel2,
Chat}. The bulk viscosity of normal conducting strange quark matter 
was also calculated \cite{Sawyer2,Madsen}. The latter might be 
relevant if the baryon density in the central regions of neutron 
stars is so high that matter becomes deconfined. 

The physical conditions in the interior of such stars are quite 
unique: this is the only place in the Universe where a deconfined 
state of cold and dense baryonic matter can naturally exist. 
This possibility has attracted a lot of attention since the 
notion of quarks was introduced \cite{Ivanenko1965,Ivanenko1969,
Itoh1970,Iachello1974,Collins1975}. 

If deconfined quark matter does exist inside stars, it is most 
likely color-superconducting. (For reviews on color 
superconductivity, see Refs.~\cite{RajWil,Alfordreview,Reddy2002,
Rischke2003,Buballa2003,Huangreview,Shovkovy2004}.)
It is therefore of great interest to study various transport 
properties of color-superconducting phases of quark matter. First 
attempts have already been made to estimate the heat and electrical
conductivity \cite{ShovEllis1,ShovEllis2}, as well as the bulk 
and shear viscosities \cite{Madsenprl2,Manuel} in the 
color-flavor-locked (CFL) phase of quark matter. Also, in the 
case of the two-flavor color-superconducting (2SC) phase, one can 
argue that most transport coefficients are dominated by the two 
ungapped (blue) quasiparticles \cite{Shovkovy2004}. (For a recent
detailed study of the bulk viscosity in the 2SC phase, see 
Ref.~\cite{Alford:2006gy}.)

In this paper, we calculate the bulk viscosity of the four most popular 
spin-one color-superconducting phases of two-flavor (non-strange) 
quark matter: the color-spin-locked (CSL), planar, polar, and the {\it A} phase 
\cite{Iwasaki,PD1,SchaferSpin1,Schmitt:2002sc,BuballaSpin1,AlfordSpin1,SchmittSpin1}. 
One of these is likely to be the ground state of dense baryon matter 
if the spin-zero Cooper pairing of quarks is prevented by the 
constraints of charge neutrality and $\beta$-equilibrium. Moreover,
cooling calculations for neutron stars favor small gaps of the 
order of $1$~MeV \cite{Grigorian:2004jq} which is the typical size of the gap in spin-one 
color superconductors \cite{Iwasaki,PD1,SchaferSpin1,Schmitt:2002sc,BuballaSpin1,AlfordSpin1,SchmittSpin1}.
The absence of strange quarks in the system may be natural if 
the medium-modified constituent value of the strange quark mass
is larger than the corresponding value of the chemical potential. 
The generalization of this study to the case of spin-one 
color-superconducting strange quark matter will be reported 
elsewhere \cite{SadShR}.

As is well known, in fully gapped spin-zero color-superconducting phases, 
the thermal densities of the quark- and hole-type quasiparticles 
are suppressed exponentially by the energy gap, 
$n_{\rm qp}\propto\exp(-\phi_{0}/T)$ where $\phi_{0}$ and $T$ 
are the values of the spin-zero gap and the temperature, respectively. 
This then translates into an exponential suppression of the 
quasiparticle contributions to the transport coefficients. 
In Ref.~\cite{Madsenprl2} such an argument was used in order to get 
a simple estimate of the viscosity in the CFL phase. It should 
be noted, however, that many transport properties are not 
dominated by the quasiparticles when there exist Nambu-Goldstone 
excitations in the low-energy spectrum, as is the case in the 
CFL phase \cite{ShovEllis1,ShovEllis2,Manuel}. The situation 
is expected to be different also in the case of spin-one phases 
which, in general, are not isotropic, and whose gap functions may 
have nodes for some directions of momenta. 

The effect of non-isotropic gaps and various topologies of nodes 
on the neutrino emission and the cooling rate of spin-one 
color-superconducting phases were recently discussed in detail 
\cite{SSW2,Wang2006}. Following a similar approach, in this work 
we study the bulk viscosity. 

Since the order parameter in spin-one color superconductors breaks
rotational invariance, 
dissipative hydrodynamics is more complicated than for isotropic
media. For instance, we expect that the viscosity becomes a
tensor \cite{prb15Bhatta}.
We shall avoid these complications by making the implicit assumption that
the hydrodynamic equations are averaged over solid angle. In this way,
only one (angular-averaged) bulk viscosity coefficient, $\zeta$, will
appear in the hydrodynamic equations 
and will have to be extracted from the (angular-averaged) neutrino emission rate.

In addition, one of the spin-one superconducting phases studied here, namely the
CSL phase, breaks baryon number, i.e., it is also a superfluid.
Fortunately, it is not an anisotropic superfluid because the order
parameter does not break rotational invariance. (Here we assume that the magnetic
field of the star is not strong enough to align the spins of the
Cooper pairs.)
Nevertheless, isotropic superfluids still have a rather complicated
hydrodynamic behavior, involving 
three (instead of only one) bulk viscosity coefficients \cite{LL6,Son:2005tj}.
However, it is not completely unrealistic to assume that the {\em
relative\/} velocity between normal and 
superfluid components is negligible compared to the {\em absolute\/} velocity of the
normal component. In this case, only one coefficient contributes to energy
dissipation. In this sense, our treatment is completely analogous to that of
Ref.\ \cite{Haensel1}, the difference being that here we consider quark
matter instead of nucleonic matter.
Let us finally note that the dissipative hydrodynamics of {\em anisotropic\/} superfluids is
even more complicated, see for instance the case of superfluid He-3 
\cite{Graham1974,prb15Bhatta}. 

In the CSL phase the 
breaking of baryon number gives rise to a phonon as the corresponding Goldstone
excitation. We neglect the contribution of the Goldstone mode to
the bulk viscosity coefficient for the following reason.
First, the effective theory for phonons is 
approximately scale-invariant at very low energies. 
For scale-invariant superfluids, however, two of the three bulk viscosity
coefficients have been shown to vanish \cite{Son:2005tj}. The remaining
coefficient may be non-zero, but corresponds to dissipation due to
relative motion of superfluid and normal component, which we have
already assumed to vanish.

The remainder of this paper is organized as follows.
In the next section, we introduce the formalism for calculating 
the bulk viscosity in non-strange quark matter. In Sec.~\ref{bulk-vis-normal}
we calculate the bulk viscosity in the normal phase of two-flavor quark 
matter. Then, in Sec.~\ref{bulk-vis-spin-1} we present our results 
for the bulk viscosity in spin-one color-superconducting phases.
The discussion of the results is given in Sec.~\ref{Conclusion}.

\section{Bulk viscosity}
\label{bulk-viscosity}

\begin{figure}
\noindent
\includegraphics[width=0.4\textwidth]{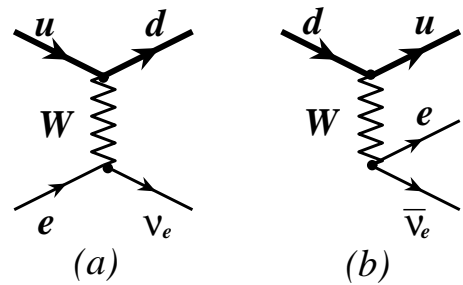}
\caption{Diagrammatic representation of the weak processes
that contribute to the bulk viscosity of non-strange quark 
matter in stellar cores.}
\label{fig-Urca_d_u_e}
\end{figure}

As mentioned in the introduction, the bulk viscosity is responsible 
for the damping of density oscillations in compact stars. The 
characteristic frequencies of interest (e.g., set by the r-mode 
instabilities) are comparable to the rotational frequencies of 
stars, i.e., $\omega\lesssim 10^{4}~\mbox{s}^{-1}$ 
\cite{Anderssonreview,Lind-lect}. These frequencies are many 
orders of magnitude smaller than the typical rates of strong 
interactions, and therefore quark matter cannot be driven 
substantially out of equilibrium with respect to strong processes. 
This is the reason why the bulk viscosity is dominated by the
much slower, flavor-changing weak processes \cite{Sawyer2,Madsen}. 
In the case of non-strange quark matter studied here, the relevant 
processes are electron capture by $u$ quarks and $\beta$ 
decay of $d$ quarks, see Fig.~\ref{fig-Urca_d_u_e}.

Let us assume that small oscillations of the quark matter density 
are described by $\delta n = \delta n_{0}\, \mbox{Re}(e^{i\omega t})$ 
where $\delta n_{0}$ is the magnitude of the density variations. 
For such a periodic process, the bulk viscosity $\zeta$ is defined 
as the coefficient in the expression for the energy-density dissipation 
averaged over one period, $\tau=2 \pi/\omega$, 
\begin {equation}
\langle \dot{\cal E}_{\rm diss}\rangle =-\frac{\zeta}{\tau} 
\int_0^{\tau} dt \left(\nabla \cdot \vec v\right)^2,
\label{epsilon-kin}
\end{equation}
where $\vec v$ is the hydrodynamic velocity associated with the density 
oscillations. By making use of the continuity equation, 
$\dot{n}+n\,\nabla\cdot\vec v=0$, we derive
\begin{equation}
\langle \dot{\cal E}_{\rm diss}\rangle 
=-\frac{\zeta \omega^2}{2}\left(\frac {\delta n_0}{n}\right)^{2}.
\label{zeta-def}
\end{equation}
In order to solve for $\zeta$, the dissipated energy on the
left-hand side has to be calculated explicitly. This can be 
done as follows.

The density oscillations drive quark matter slightly out of 
$\beta$ equilibrium, but not out of thermal equilibrium which 
is restored almost without delay by strong processes. The 
corresponding instantaneous quasi-equilibrium state can be 
unambiguously characterized by the total baryon number density
$n$ and the lepton fraction $X_e$,
\begin{subequations}
\begin{eqnarray}
n &=& \frac13\left(n_u+n_d\right), \\
X_e &=& \frac{n_e}{n},
\end{eqnarray}
\label{composiiton}
\end{subequations}
where $n_u$ and $n_d$ are the number densities of up and down quarks,
while $n_e$ is the number density of electrons. (In the case of strange 
quark matter, one should also add the strangeness fraction $X_s=n_s/n$ 
where $n_s$ is the number density of strange quarks \cite{SadShR}.) 
Charge neutrality requires
\begin{equation}
\frac23 n_u-\frac13 n_d-n_e=0. 
\label{neutral}
\end{equation}
Using this constraint together with the definitions in Eq.~(\ref{composiiton}),
one can express the number densities and, in fact, all thermodynamic quantities 
of quark matter in terms of $n$ and $X_e$. For the number densities, for example, 
one finds
\begin{subequations}
\begin{eqnarray}
n_e &=& X_e n,   \label{rho_e}\\
n_u &=&(1+X_e) n,   \label{rho_u}\\
n_d &=&(2-X_e) n.  \label{rho_d}
\end{eqnarray}
\end{subequations}
These number densities can also be expressed in terms of the corresponding 
chemical potentials, $n_i=n_i(\mu_i)$. In $\beta$ equilibrium, the three 
chemical potentials are related as follows: $\mu_d=\mu_u+\mu_e$. In pulsating 
matter, on the other hand, the instantaneous departure from equilibrium is
described by the small parameter
\begin{equation}
\delta\mu \equiv \mu_d-\mu_u-\mu_e =  \delta\mu_d-\delta\mu_u-\delta\mu_e,
\label{delta-mu-def}
\end{equation}
where $\delta\mu_{i}$ denotes the deviation of chemical potential $\mu_i$
from its value in $\beta$ equilibrium. The quantity $\delta\mu$ 
can be conveniently expressed in terms of the variations of 
the two independent variables $\delta n$ and $\delta X_e$, 
\begin{equation}
\delta\mu = C\frac{\delta n}{n} + B \delta X_e,
\label{delta-mu}
\end{equation}
where, as follows from the definition, the coefficient functions $C$ 
and $B $ are given by 
\begin{subequations}
\begin{eqnarray}
C &=& n_d \frac{\partial \mu_d}{\partial n_d}
     -n_u \frac{\partial \mu_u}{\partial n_u}
     -n_e \frac{\partial \mu_e}{\partial n_e},   \label{def-C}\\
B  &=& -n \left(\frac{\partial \mu_d}{\partial n_d} 
                 + \frac{\partial \mu_u}{\partial n_u}
                 + \frac{\partial \mu_e}{\partial n_e} \right).
\label{def-B_e}
\end{eqnarray}
\label{def-C-B_e}
\end{subequations}
When $\delta\mu$ is non-zero the two Urca processes, shown diagrammatically 
in Fig.~\ref{fig-Urca_d_u_e}, have slightly different rates. To leading order
in $\delta\mu$, we could write
\begin{equation}
\Gamma_{\nu} - \Gamma_{\bar\nu} = - {\lambda} \delta\mu .
\label{ratediff0}
\end{equation}
(Note that our $\lambda$ is defined so that it is non-negative.)
The net effect of having different rates for the two processes is
a change of the electron fraction in the system:
\begin{equation}
n \frac{d (\delta X_e)}{dt} = {\lambda} \delta\mu ,
\label{ratediff}
\end{equation}
This has the tendency to restore the equilibrium value of $X_{e}$. Since the rate is
finite, however, the weak processes always lag behind the density oscillations. 
In order to see this explicitly, we substitute $\delta\mu$ from Eq.~(\ref{delta-mu})
into Eq.~(\ref{ratediff}) and get the equation for $\delta X_e$ in a 
closed form,
\begin{equation}
n \frac{d (\delta X_e)}{dt} = {\lambda} 
\left(C\frac{\delta n}{n} + B \delta X_e\right) .
\label{eqX_e}
\end{equation}
The periodic solution to this equation can be found most easily by making 
use of complex variables. Denoting $\delta X_e \equiv \mbox{Re}
\left(\delta X_{e,0}\, e^{i\omega t}\right)$, we derive the following 
result:
\begin{equation}
\delta X_{e,0} = \frac{\delta n_0}{n}\frac{C}{i\alpha-B },
\label{deltaX_e0}
\end{equation}
where, by definition, $\alpha\equiv n \omega/\lambda$. In the last equation,
the lagging of the weak processes is indicated by a non-vanishing imaginary part
of $\delta X_{e,0}$. Such an imaginary part controls the phase shift of the 
$\delta X_e$ oscillations with respect to the oscillations of density. 

As we show in a moment, the same phase shift also leads to a non-vanishing 
dissipation of the energy density,
\begin{equation}
\langle \dot{\cal E}_{\rm diss}\rangle 
= \frac{n}{\tau} \int_0^{\tau} P \dot{V} dt
\label{diss-energy}
\end{equation}
where $V\equiv 1/n$ is the specific volume. 

The pressure oscillations around the equilibrium value are driven by the
oscillations of its two independent variables, i.e., the quark number 
density and the lepton fraction,  
\begin{equation}
\delta P = \frac{\partial P}{\partial n}\, \delta n
-n\, C\, \delta X_e ,
\label{press}
\end{equation}
where $C$ is the same as in Eq.~(\ref{def-C}). In the derivation we took 
into account that $n_i =\partial P/\partial \mu_i$ and that the total 
pressure is given by the sum of the partial contributions of the quarks 
and electrons, $P=\sum_{i}P_{i}(\mu_i)$.

After taking into account the relation (\ref{press}) together with the 
solution for $\delta X_{e,0}$ in Eq.~(\ref{deltaX_e0}), the expression
(\ref{diss-energy}) becomes
\begin{eqnarray}
\langle \dot{\cal E}_{\rm diss}\rangle 
&=& \frac{C}{2} \omega\, \delta n_0 \, \mbox{Im}\left(\delta X_{e,0}\right)
\nonumber \\
&=& -\frac{1}{2}\left(\frac{\delta n_0}{n}\right)^2\frac{\lambda\, \omega^2 C^2}
{\omega^2+\left(\lambda B /n\right)^2}.
\label{diss}
\end{eqnarray}
By comparing this with the definition in Eq.~(\ref{zeta-def}), we finally derive 
an explicit expression for the bulk viscosity,
\begin{equation}
\zeta = \frac{\lambda C^2}{\omega^2+\left(\lambda B /n\right)^2} .
\end{equation}
This expression shows that the viscosity is maximum in the limit of zero
frequency, $\zeta_{\rm max }=\zeta_{\omega=0}$, and that it falls off as 
$1/\omega^2$ at high frequencies, $\omega\gg \omega_0\equiv \lambda B /n$. 
It should be also noted that $\omega_{0}\sim \lambda$, and that the maximum 
viscosity is inversely proportional to the rate, i.e., $\zeta_{\rm max}
\sim 1/\lambda$. Because the rates of the weak processes in a dense medium 
usually have a power-law (or even an exponential) dependence on the 
temperature, the bulk viscosity is a very sensitive function of the 
temperature, too.

\section{Bulk viscosity in the normal phase}
\label{bulk-vis-normal}

In order to calculate the bulk viscosity in the normal phase of two-flavor 
quark matter, we need to determine the corresponding thermodynamic coefficients 
$B $ and $C$ [see Eq.~(\ref{def-C-B_e})] and calculate the difference of the 
rates of the two Urca processes shown in Fig.~\ref{fig-Urca_d_u_e}. 


By making use of the following relations valid for non-interacting 
quark matter
\begin{subequations}
\begin{eqnarray}
n_{u,d} &=& \frac{1}{\pi^2}\left(\mu_{u,d}^2-m_{u,d}^2\right)^{3/2}, \\
n_{e}&=& \frac{1}{3\pi^2}\mu_e^3,
\end{eqnarray}
\end{subequations}
we derive
\begin{subequations}
\begin{eqnarray}
C &\simeq & \frac{m_u^2}{3\mu_u}-\frac{m_d^2}{3\mu_d} ,
\label{C-free-quark}\\
B &\simeq & \frac{\pi^2}{3}n\left(\frac{1}{\mu_u^2}
+\frac{1}{\mu_d^2}+\frac{3}{\mu_e^2}\right).
\label{B-free-quark}
\end{eqnarray}
\label{C-B-normal}
\end{subequations}
Here we made use of the equilibrium relation satisfied by the chemical
potentials, $\mu_d=\mu_u+\mu_e$, and neglected higher-order mass 
corrections in the expression for $B $. In the temperature regime of 
interest, $T\lesssim m_{u,d}$, we also do not need to take into 
account any corrections due to a non-zero temperature. 

It should be noted that the coefficient $C$, and therefore the bulk 
viscosity which is proportional to $C^2$, vanishes in the case of 
massless quarks. Moreover, this statement remains true even if the 
following (non-)Fermi liquid correction due to strong forces are 
taken into account \cite{Freedman1976,Baluni1977,Fraga2004,SchaferChapter},
\begin{equation}
C^{\prime} = \frac{4\alpha_s}{3\pi} \left[\frac{m_d^2}{\mu_d}
\left(\ln\frac{2\mu_d}{m_d}-\frac23\right)
-\frac{m_u^2}{\mu_u}
\left(\ln\frac{2\mu_u}{m_u}-\frac23\right)\right].
\label{C-prime}
\end{equation}
(For a recent discussion of (non-)Fermi liquid corrections see, 
for example, Ref.~\cite{SchaferChapter}.) Because of the large 
logarithms on the right-hand side of Eq.~(\ref{C-prime}), this 
correction can become even larger than the leading-order result 
for $C$ in Eq.~(\ref{C-free-quark}). Our estimates show that, in 
two-flavor quark matter, taking $C^{\prime}$ into account may 
increase the value of the viscosity by approximately an order 
of magnitude. Therefore, in all numerical estimates below we 
add the contributions from Eqs.~(\ref{C-free-quark}) and 
(\ref{C-prime}).

\begin{figure}
\includegraphics[width=0.48\textwidth]{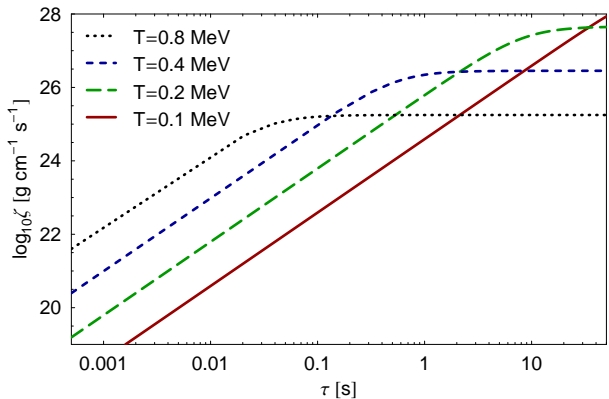}
\caption{The bulk viscosity for the normal phase of two-flavor 
quark matter as a function of the period of the density oscillations.}
\label{bv-fig1}
\end{figure}

Now let us turn to the calculation of $\lambda$ defined by 
Eq.~(\ref{ratediff0}). Following the original approach of Iwamoto 
\cite{Iwamoto}, we get the rate for $\beta$ decay in the following
form:
\begin{widetext}
\begin{equation}
\Gamma_{\bar \nu}(\delta \mu)= 
6\int \frac{d^3 p_d d^3 p_u d^3 p_{\bar\nu} d^3 p_e}{(2\pi)^{8} E_d E_u E_{\bar\nu} E_e}
\vert M \vert ^2 \delta ^4 \left(P_d-P_u-P_e-P_{\bar \nu}\right)
f\left({E_d-\mu _d}\right)\left[ 1-f\left({E_u-\mu _u}\right)
\right] \left[ 1-f\left({E_e-\mu _e}\right)\right].
\label{Gamma}
\end{equation}
\end{widetext}
Here, $P_i$ and $p_i$ are the 4- and 3-momenta of the $i$th particle, 
respectively, 
and $f(E)\equiv 1/(e^{E/T}+1)$ is the Fermi distribution function.
The scattering amplitude squared is given by \cite{Iwamoto}
\begin{eqnarray}
\vert M \vert ^2 &=& 64 G_F^2 \cos^2\theta_C(P_d\cdot P_{\bar \nu})(P_u \cdot P_e)\nonumber\\
&\approx&  \frac{2^8\alpha _s}{3 \pi} G_F^2 \cos^2\theta_C E_u E_d E_{\bar\nu} E_e 
(1-\cos\theta_{d\bar\nu}).
\end{eqnarray}
After substituting this approximate form of $\vert M \vert ^2$, all 
angular integrals in Eq.~(\ref{Gamma}) can be done exactly. Then, using 
the dimensionless variables $x_i=({E_i - \mu_i})/{T}$ (note that 
$\mu_{\bar\nu}=0$), we obtain
\begin{equation}
\Gamma_{\bar\nu}(\xi)=\frac{4\alpha _s}{\pi^6} G_F^2\cos^2 \theta_C 
\mu_d \mu_u \mu_e T^5  
\int_0^{\infty}dx_{\bar\nu}x_{\bar\nu}^2 J(x_{\bar\nu}-\xi),
\end{equation}
where $\xi \equiv {\delta \mu}/{T}$, and
\begin{eqnarray}
J(x)&=&\left[ \prod_{j=1}^{3} \int_{-\infty}^{\infty}dx_j f(x_j)\right] 
\delta(x_1+x_2+x_3-x)\nonumber\\
&=& \frac{\pi^2+x^2}{2(1+e^x)}.
\end{eqnarray}
By noting that 
$\Gamma_{\bar\nu}(\xi)=\Gamma_{\nu}(-\xi)$, we finally derive the expression 
\begin{eqnarray}
\lambda &=& \frac{17}{15\pi^2} G_F^2\cos^2 \theta_C \alpha _s \mu_d \mu_u \mu_e T^4 
\nonumber\\
&\simeq & \frac{17}{15} G_F^2\cos^2 \theta_C \alpha _s (6X_{e})^{1/3} n\, T^4 \left[1+{\cal O}(X_e)\right].
\label{lambda}
\end{eqnarray}
This result, together with the explicit form for the coefficient functions 
$C$, $C^{\prime}$, and $B $ in Eqs.~(\ref{C-B-normal}) 
and (\ref{C-prime}), is sufficient to calculate the bulk 
viscosity of the normal phase of dense quark matter. The numerical results 
are presented in Fig.~\ref{bv-fig1}. In the calculation, we used the following
representative values of the parameters:
\begin{subequations}
\begin{eqnarray}
\mu_d &=& 500~\mbox{MeV}, \quad m_d=9~\mbox{MeV},\\
\mu_u &=& 400~\mbox{MeV}, \quad m_u=5~\mbox{MeV},\\
\mu_e &=& 100~\mbox{MeV},  \quad \alpha_s = 1.
\end{eqnarray}
\label{parameters}
\end{subequations}
In order to understand the numerical results better, we calculate
the maximum value of the bulk viscosity, $\zeta_{\rm max} \equiv 
\zeta(\omega=0)$, and the characteristic frequency $\omega_0= 
\lambda B /n$, which separates the low- and high-frequency regimes, 
\begin{eqnarray}
\zeta_{\rm max} &\approx & 7.26\times 10^{24}
\left(\frac{T}{1~\mbox{MeV}}\right)^{-4}
~\mbox{g}~\mbox{cm}^{-1}\mbox{s}^{-1} , \label{MAXzeta} \\
\omega_0 &\approx & 
460 \left(\frac{T}{1~\mbox{MeV}}\right)^{4}~\mbox{s}^{-1} .
\label{omega0}
\end{eqnarray}
The knowledge of these two quantities is sufficient for 
determining the bulk viscosity at all frequencies, 
\begin{equation}
\zeta = \frac{\zeta_{\rm max}}{1+(\omega/\omega_0)^2} .
\label{general-form}
\end{equation}
Taking into account the relations in Eqs.~(\ref{MAXzeta}) and 
(\ref{omega0}), the temperature dependence of the results in 
Fig.~\ref{bv-fig1} becomes clear. In accordance with
the first relation, the maximum value of the viscosity (approached 
at low frequencies or large $\tau$) becomes larger with decreasing 
temperature. Also, in agreement with Eq.~(\ref{omega0}), the 
location of the ``shoulder" (occurring at $\tau_0\equiv 2\pi/\omega_0$) 
shifts to smaller periods (higher frequencies) when the temperature 
gets higher.

\section{Bulk viscosity in spin-one color-superconducting phases}
\label{bulk-vis-spin-1}

In this section, we calculate the bulk viscosity of several spin-one 
color-superconducting phases that have been proposed and studied in
detail in Refs.~\cite{PD1,SchaferSpin1,BuballaSpin1,AlfordSpin1,SchmittSpin1}.
Following Ref.~\cite{SSW2}, we focus on the so-called transverse 
phases in which only quarks of opposite chiralities pair. Theoretical 
studies at asymptotically large densities suggest that such phases are 
preferred \cite{SchaferSpin1,SchmittSpin1}. 

In $\beta$ equilibrium, the rates of the Urca processes were calculated 
in Ref.~\cite{SSW2} in four different spin-one phases. Following the same
method, here we generalize the calculation to the quasi-equilibrium state, 
characterized by a small but non-vanishing value of $\delta\mu=\mu_d-\mu_u-\mu_e$.

Starting from equations analogous to Eq.~(36) and (37) in Ref.~\cite{SSW2}, 
we derive
\begin{widetext}
\begin{eqnarray}
\Gamma_{\nu}-\Gamma_{\bar\nu} &=&\frac{2}{3 \pi^6}\alpha_s G^2 \mu_e \mu_u \mu_d T^5 
\sum _{r}\int _{-1}^{1}d\, \Xi \int x_{\nu}^{2} dx_{\nu}
 \left[ F^{rr}_{\varphi _u \varphi _d} 
(\Xi,x_\nu+\xi)-F^{rr}_{\varphi _u \varphi _d} (\Xi,x_\nu-\xi)\right],
\label{rate-sp1}
\end{eqnarray}
where $\Xi \equiv \cos\theta_u=\cos\theta_d$ is the angle 
between the three-momentum of the $u$/$d$ quark and the $z$-axis.
The functions $F^{rr}_{\varphi _u \varphi _d} (\Xi,x)$ are the same 
as in Ref.~\cite{SSW2}. Their explicit form is given by
\begin{eqnarray}
F^{rr}_{\varphi _u \varphi _d} (\Xi,x) &=& \omega_{rr}(\Xi)
\sum_{e_1,e_2=\pm}\int_{0}^{\infty}\! \int_{0}^{\infty} dx_d dx_u 
\left(e^{-e_{1}\sqrt{x_u^2+\lambda_{\Xi,r}\varphi_u^2}}+1\right)^{-1}
\left(e^{e_{2}\sqrt{x_d^2+\lambda_{\Xi,r}\varphi_d^2}}+1\right)^{-1} 
\nonumber\\
&&\times \left(e^{x_\nu+e_{1}\sqrt{x_u^2+\lambda_{\Xi,r}\varphi_u^2}
-e_{2}\sqrt{x_d^2+\lambda_{\Xi,r}\varphi_d^2}}+1\right)^{-1},
\end{eqnarray}
\end{widetext}
where, by definition, $\xi=\delta\mu/T$ and $\varphi_i = {\phi_i}/{T}$. 
For an explicit form of the functions $\lambda_{\Xi,r}$ and $\omega_{rr}$, 
we refer the reader to Refs.~\cite{SchmittSpin1,SSW2}. 

\begin{figure}
\includegraphics[width=0.48\textwidth]{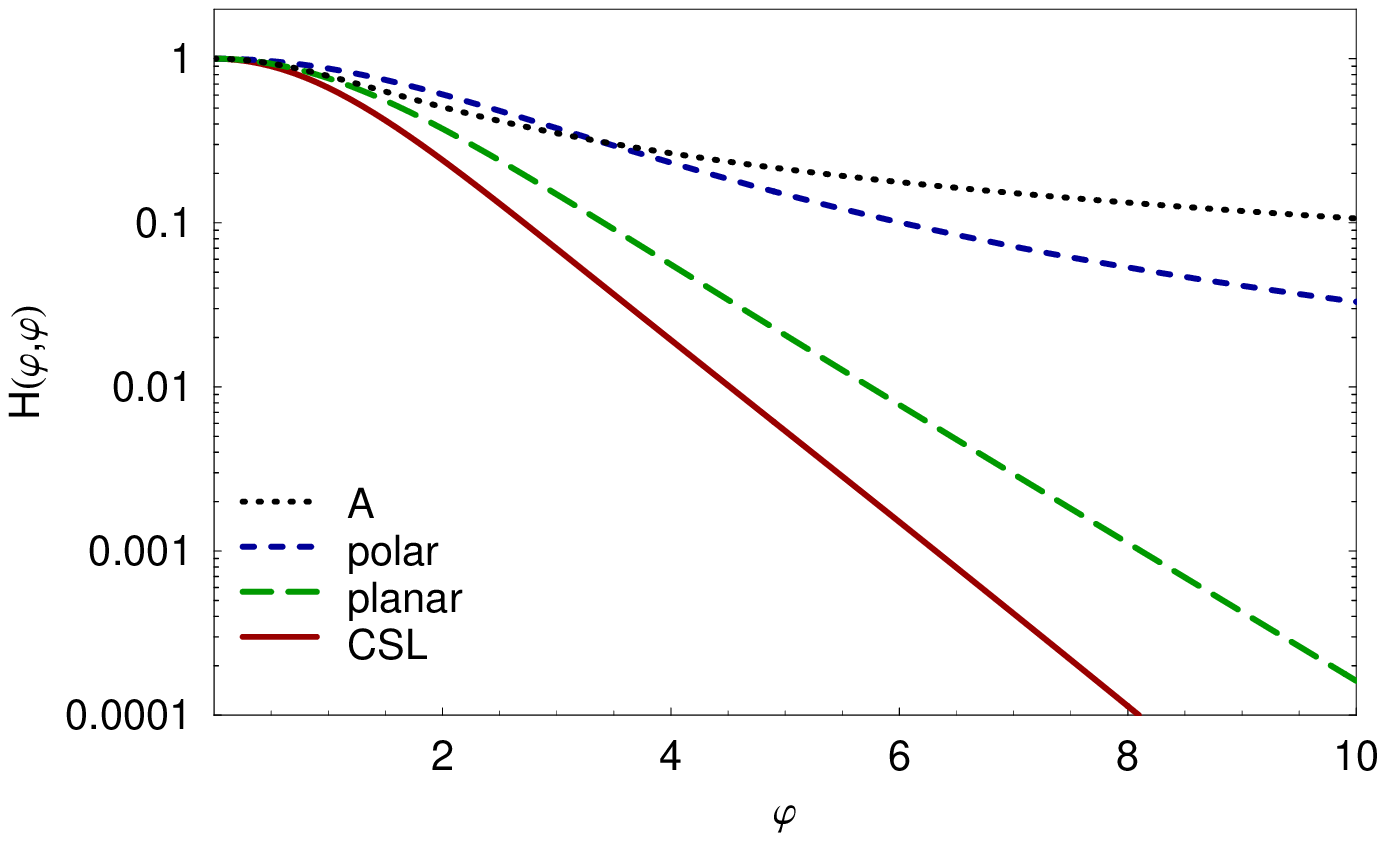}
\caption{The reduction factor as a function of $\varphi\equiv \phi/T$ 
for the CSL, planar, polar, and {\it A} phases.}
\label{bv-fig2}
\end{figure}

By expanding the rate difference (\ref{rate-sp1}) in powers of $\xi$ and 
extracting the coefficient of the linear term, we derive an expression 
for $\lambda$ in the following general form:
\begin{equation}
\lambda(\varphi_d,\varphi_u)=\lambda(0)
\left[\frac{1}{3}+\frac{2}{3}H(\varphi_d,\varphi_u)\right] ,
\label{def-lambda-rate}
\end{equation}
where $\lambda(0)$ is the same as in the normal phase of quark matter,
see Eq.~(\ref{lambda}), and $H(\varphi_d,\varphi_u)$ is a reduction 
factor whose explicit form depends on the choice of a specific spin-one 
color-superconducting phase. Note that, as in the case of the 
neutrino-emission rates \cite{SSW2,Wang2006}, $\lambda$ consists of two 
qualitatively different contributions. The first one is given by the 
term $1/3$ in the square brackets of Eq.~(\ref{def-lambda-rate}). It 
originates from the ungapped modes that are present in all considered 
spin-one phases. The second contribution is given by the term proportional 
to $H(\varphi_u,\varphi_d)$. This one originates from the gapped modes.
An explicit form of the function $H(\varphi_u,\varphi_d)$ in the case 
of the CSL, planar, and polar phases is given by
\begin{equation}
H(\varphi_d,\varphi_u) = \frac{60}{17\pi^4}\int_{-1}^1 d\,\Xi
\int_0^{\infty} dx_{\nu} x_{\nu}F^{11}_{\varphi _u \varphi _d} (\Xi,x_\nu),
\end{equation}
and, in the case of the {\it A} phase, it reads
\begin{eqnarray}
 H(\varphi_d,\varphi_u) &=& \frac{60}{17\pi^4}
\int_{-1}^1 d\,\Xi \int_0^{\infty} dx_{\nu} x_{\nu} \nonumber\\
&\times&\left[ F^{11}_{\varphi_u \varphi_d} (\Xi,x_\nu)
-F^{22}_{\varphi _u \varphi _d} (\Xi,x_\nu)\right] .
\end{eqnarray}
In the derivation we integrated by parts so that the results are given in 
terms of the function $F^{rr}_{\varphi _u \varphi _d} (\Xi,x)$.
The reduction factors can be easily calculated numerically for each 
phase. The results are shown in Fig.~\ref{bv-fig2} for 
the case $\varphi_u=\varphi_d \equiv\varphi$. In the limit of 
large $\varphi$ analytical expressions for the suppression 
functions for all four spin-one color-superconducting phases are 
given in Appendix~\ref{App}.

\begin{figure}
\includegraphics[width=0.48\textwidth]{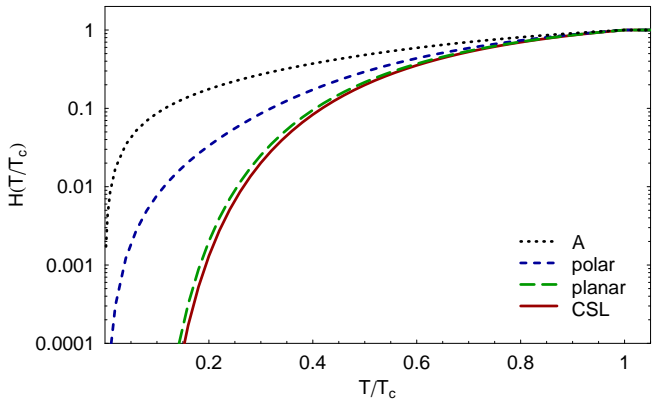}
\caption{The reduction factor as a function of the temperature
for the CSL, planar, polar, and {\it A} phases.}
\label{bv-fig3}
\end{figure}

For many applications, it is of interest to know the temperature 
dependence of the rates. Therefore, in Fig.~\ref{bv-fig3}, we also 
show the temperature dependence of the suppression factor $H(T)$. 
In the calculation we used the following temperature dependence 
of the gap parameter, 
\begin{equation}
\phi(T)=\phi_{0}\,\sqrt{1-\left(\frac{T}{T_c}\right)^2},
\label{Phi_of_T}
\end{equation}
with $\phi_{0}$ being the value of the gap parameter at $T=0$, and $T_c$ 
being the value of the critical temperature. Note that the ratio 
$T_c/\phi_{0}$ depends on the choice of the phase \cite{SchmittSpin1}.
The approximate values of this ratio are $0.8$ (CSL), $0.66$ (planar),
$0.49$ (polar), and $0.81$ ({\it A} phase).

With the expression for $\lambda$ at hand, we are now in the position to 
calculate the bulk viscosity of the spin-one color-superconducting 
phases. Before we do this, it might be appropriate to mention that 
the coefficient functions $B $ and $C$ do not change much due to 
superconductivity. Corrections to $B $ are of order 
$(\phi_{i}/\mu_{i})^2$ and, thus, are strongly suppressed.
The function $C$, on the other hand, gets corrections
of order $\phi_i^2/\mu_i$ which are 
negligible only if $\phi_{i}^2\ll m_{i}^2$. We assume that this 
is indeed the case. Notably, the spin-one gap corrections to $C$ 
should be comparable to the corrections due to a non-zero temperature 
that we neglected in our calculations. 

It should be emphasized that the bulk viscosity in the color-superconducting 
phases has the same general structure as in the normal phase, see 
Eq.~(\ref{general-form}), but the quantities $\zeta_{\rm max}$ and 
$\omega_0$ should be redefined to take into account the rate suppression 
factors:
\begin{eqnarray} 
\zeta_{\rm max}^{\rm sp1} &=& \frac{\zeta_{\rm max}}{h_{\rm sp1}}, \\
\omega_0^{\rm sp1} &=& h_{\rm sp1} \omega_0,
\end{eqnarray}
where $\zeta_{max}$ and $\omega_0$ are given in Eqs.~(\ref{MAXzeta}) 
and (\ref{omega0}), respectively, and
\begin{equation}
h_{\rm sp1} = \frac13+\frac23 H(T/T_c). \label{h-sp1}
\end{equation}
We see that the effect of the suppression of $\lambda$ due to 
Cooper pairing manifests itself in a non-trivial way in the expression 
for the bulk viscosity, i.e.,
\begin{equation}
\zeta_{\rm sp1} = \frac{ \zeta_{\rm max}h_{\rm sp1}}
{h_{\rm sp1}^2 +(\omega/\omega_0)^2}.
\label{zeta-sp1}
\end{equation}
{}From the analysis of this representation, we find that the 
suppression of the rates tends to decrease the viscosity at high
frequencies ($\omega>h_{\rm sp1}\omega_0$) and to increase it at 
low frequencies ($\omega<h_{\rm sp1}\omega_0$). One should
note, though, that the relevant range of low frequencies would 
shrink a lot if the suppression happened to be strong. 

The representation in Eq.~(\ref{zeta-sp1}) shows that the effect 
of color superconductivity cannot be very large. Even if the 
suppression of the weak rates due to gapped modes is maximal, 
i.e., $H(T/T_c)=0$, the results for $\zeta_{\rm max}$ and 
$\omega_0$ could not change by more than a factor of $3$
compared to the normal phase of matter. Of course, this is the 
consequence of having ungapped quasiparticle modes in the energy 
spectra. 

\begin{figure}
\includegraphics[width=0.48\textwidth]{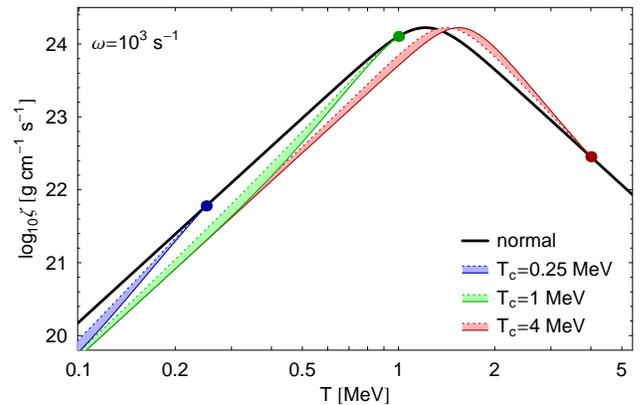}
\caption{The bulk viscosity as a function of the temperature for 
spin-one color superconducting quark matter. The oscillation 
frequency is $\omega=10^3~\mbox{s}^{-1}$.}
\label{bv-fig4}
\end{figure}

The numerical results for the bulk viscosity in the normal phase and 
spin-one color-superconducting phases with different values of the 
critical temperature are shown in Fig.~\ref{bv-fig4}. For each of the 
three choices of $T_c$, we show a shaded area (color online: blue, 
green and red for $T_c=0.25$, $1$, and $4$~MeV, respectively), that 
is bounded by the values of the bulk viscosity in the CSL phase (solid 
boundary) and in the {\it A} phase (dotted boundary). In the calculation, 
we used the set of parameters given in Eq.~(\ref{parameters}).

\begin{figure}
\includegraphics[width=0.48\textwidth]{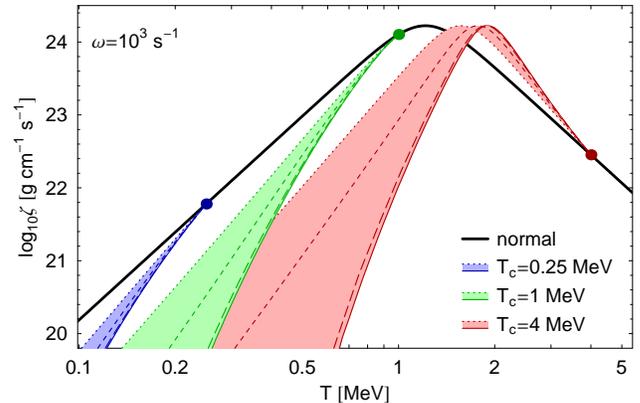}
\caption{The bulk viscosity as a function of the temperature for 
a toy model of a spin-one color superconductor in which all 
quasiparticle modes are gapped. The oscillation frequency is 
$\omega=10^3~\mbox{s}^{-1}$.}
\label{bv-fig5}
\end{figure}

For comparison, in Fig.~\ref{bv-fig5} we also show the results 
for the bulk viscosity in a toy model in which all quasiparticles 
modes are gapped. In this case, the effect is indeed very dramatic.
The largest deviations from the result of the normal phase correspond 
to the CSL phase (thin solid lines), and the smallest to the 
{\it A} phase (dotted lines). The viscosity in the planar phase 
(long-dashed lines) is numerically close to that in the CSL phase,
and the viscosity in the polar phase (short-dashed lines) is 
typically somewhere in between. 

At this point it is natural to ask ourselves if there exists 
the possibility that all quasiparticles in a spin-one color 
superconductor are gapped. The answer is affirmative. This 
is the case, e.g., in the version of the CSL phase proposed 
in Ref.~\cite{ABBY}, when the quarks are massive. In our notation, 
the three quasiparticle gaps are given by
\begin{equation}
\phi_{i,1}=\phi_{i,2}\simeq\phi_{i}, \quad \phi_{i,3} 
\simeq \frac{m_i\phi_i}{\sqrt{2}\mu_i},
\quad (i=u,d).
\end{equation}
The qualitative change in the 
low-energy spectrum comes from a different choice of the gap 
matrix in the CSL phase. (For a discussion of the specific 
differences, see Sec.~VII in Ref.~\cite{SSW2}.) In fact, it 
is likely that the ansatz that produces the fully gapped phase
is energetically favored \cite{BuballaComm}. The most important 
effect of the additional non-zero gap is the appearance of 
another suppression factor multiplying the first term in the 
expression for $\lambda$, see Eq.~(\ref{def-lambda-rate}). 
Therefore, the modified expression for the CSL phase reads
\begin{equation}
\lambda(\varphi_d,\varphi_u)=\lambda(0)
\left[\frac{1}{3}H\left(
\frac{m_d \varphi_d}{\sqrt{2}\mu_d},
\frac{m_u \varphi_u}{\sqrt{2}\mu_u}\right)
+\frac{2}{3}H(\varphi_d,\varphi_u)\right] .
\label{lambda-rate}
\end{equation}
Note that the additional suppression factor is given in terms 
of the same function $H(\varphi_d,\varphi_u)$ which was calculated 
numerically for the transverse CSL phase, see 
Figs.~\ref{bv-fig2} and \ref{bv-fig3}. If the up and down quark 
masses are rather small, the effect from the additional suppression 
cannot be easily seen before the temperature becomes much lower than the 
critical value, i.e., $T\ll T_c$. The situation changes dramatically, 
however, if the relevant constituent values of the quark masses
happen to be considerably larger than the current masses of quarks.
As suggested by the analysis in Ref.~\cite{ABBY}, this is indeed 
possible. Then, the bulk viscosity could be affected almost as much 
as in the toy model in Fig.~\ref{bv-fig5}.

\section{Discussion}
\label{Conclusion}

In this paper we have calculated the bulk viscosity for the normal
phase as well as for four spin-one color-superconducting phases of 
two-flavor dense quark matter. The main contributions come from 
the Urca processes shown diagrammatically in 
Fig.~\ref{fig-Urca_d_u_e}. Note that the results for the normal 
phase are also relevant for the 2SC phase. Indeed, after taking into 
account that there are two (blue) ungapped modes of quasiparticles 
in the low-energy spectrum of the 2SC phase, the low-temperature bulk 
viscosity is approximately given by the same expression (\ref{general-form}), 
provided the following redefinitions are made: $\zeta_{max}^{\rm 2SC}
=3\zeta_{max}$ and $\omega_0^{\rm 2SC}=\omega_0/3$, where the normal-phase 
quantities are given in Eqs.~(\ref{MAXzeta}) and (\ref{omega0}), 
respectively. The redefinitions account for the decrease of the weak 
rates by a factor of 3 at $T\ll \Delta_0$ where $\Delta_0$ is the value 
of the 2SC gap.

The microscopic calculations of the bulk viscosity in the spin-one 
color-superconducting phases suggests that quasiparticles with 
different types of gapless nodes (e.g., points or lines at the Fermi
sphere) could potentially play a very important role. In the case of
the transverse phases, however, the presence of a single ungapped 
quasiparticle mode washes out essentially all information about 
spin-one Cooper pairing, see Fig.~\ref{bv-fig4}. The presence of
non-zero quark masses may provide a gap for such a mode and the 
situation changes. In this paper, we briefly discussed such 
a possibility in connection with the CSL phase of Ref.~\cite{ABBY}. 
The results are shown in Fig.~\ref{bv-fig5}.

In agreement with the general expectation, we find that the bulk viscosity 
often tends to decrease when there is Cooper pairing of quarks whose main 
effect is to suppress the rates of the weak processes. In some cases (e.g.,
at sufficiently low frequencies and/or at temperatures close to the critical
value) the behavior may reverse because of the non-trivial dependence
of the bulk viscosity on the suppression factor, see Eq.~(\ref{zeta-sp1}).
Such an increase of the viscosity in the color-superconducting CSL phase 
is seen, for example, in a range of temperatures below $T_c$ in 
Figs.~\ref{bv-fig4} and \ref{bv-fig5} in the case when $T_c=4$~MeV.

\section*{Acknowledgments}

The authors acknowledge discussions with M.~Alford, D.~Bandyopadhyay, J.~Berges, 
M.~Buballa, F.~P.~Esposito, A.~Schmitt, and J.~Wambach. B.A.S. acknowledges 
support from the Frankfurt International Graduate School for Science (FIGSS). 
The work of D.H.R. and I.A.S. was supported in part by the Virtual 
Institute of the Helmholtz Association under grant No. VH-VI-041, 
by the Gesellschaft f\"{u}r Schwerionenforschung (GSI), and by the 
Deutsche Forschungsgemeinschaft (DFG).

\appendix
\section{Asymptotic form of the suppression function $H(\varphi,\varphi)$ 
at $\varphi\to\infty$}
\label{App}

In this appendix, we calculate the asymptotic form of the suppression 
function $H(\varphi,\varphi)$ at $\varphi\to\infty$. This is relevant 
for the bulk viscosity of spin-one color superconductors in the 
low-temperature regime. 

Let us start from the CSL and planar phases whose gap functions have 
no nodes in momentum space. To this end, it is useful to compute the 
asymptotic behavior of the following integral:
\begin{widetext}
\begin{eqnarray}
J(\varphi,\varphi) &\equiv & \sum_{e_1,e_2=\pm}
\int_0^\infty\! \int_0^\infty\! \int_0^\infty dx_{u} dx_{d} dx_{\nu} x_{\nu} 
\left(e^{-e_1\sqrt{x_{d}^2+\varphi^2}} + 1\right)^{-1}
\left(e^{e_2\sqrt{x_{u}^2+\varphi^2}} + 1\right)^{-1} \nonumber\\
& & \times \left(e^{x_{\nu}+e_1\sqrt{x_{u}^2+\varphi^2}
-e_2\sqrt{x_{d}^2+\varphi^2}} + 1\right)^{-1} .
\end{eqnarray}
\end{widetext}
By neglecting the terms of order ${\cal O}(e^{-2\varphi})$, we arrive at the 
following leading-order asymptotic behavior:
\begin{eqnarray}
J(\varphi,\varphi)&\simeq& 
4\varphi e^{-\varphi}\int_0^\infty\! \int_0^\infty\! \int_0^\infty 
\frac{dy_{u} dy_{d} dx_{\nu} x_{\nu}}
{e^{y_{u}^2}+e^{x_{\nu}+y_{d}^2}}\nonumber\\
&\approx & 5.047 \varphi e^{-\varphi}.
\end{eqnarray}
Note that in the derivation we changed the integration variables,
$x_{i}=y_{i}\sqrt{2\varphi}$ for $i=u,d$, and calculated the 
remaining integral numerically.

By making use of the result for $J(\varphi,\varphi)$, we 
derive the large $\varphi$ asymptotic behavior for the suppression function 
$H(\varphi,\varphi)$ in the CSL phase:
\begin{equation}
H^{\rm CSL}(\varphi,\varphi) = \frac{240}{17\pi^4} J(\sqrt{2}\varphi,\sqrt{2}\varphi) 
\simeq 1.034\varphi e^{-\sqrt{2}\varphi}.
\end{equation}
Similarly, after taking into account the angular dependence of the gap
in the planar phase, we derive
\begin{eqnarray}
H^{\rm planar}(\varphi,\varphi) &=& \frac{120}{17\pi^4} \int_{-1}^{1}d\, \Xi
J(\sqrt{1+\Xi^2}\varphi,\sqrt{1+\Xi^2}\varphi) \nonumber\\
&\simeq & 0.917\sqrt{\varphi} e^{-\varphi}.
\end{eqnarray}
The derivation in the polar and {\it A} phases is slightly more complicated
because the corresponding gap functions have zeros for some directions in 
momentum space. By approximating the angular integrals in the same way as 
in Ref.~\cite{SSW2} (see Appendix~E there), we arrive at the following
asymptotic behaviors:
\begin{eqnarray}
H^{\rm polar}(\varphi,\varphi) &\simeq& \frac{\pi}{\varphi^2},\\
H^{\rm A}(\varphi,\varphi) &\simeq & \frac{1}{\varphi}.
\end{eqnarray}
In connection to these last two results, we should mention that while the 
parametric dependence on $\varphi$ is easy to extract analytically, it is 
much harder to determine the overall coefficients in their power-law asymptotic behavior. 
In our derivation, therefore, we combined the analytical derivation with the
numerical calculations.

\bibliography{bulk_vis_nonstr_v6}

\begin{thebibliography}{59}
\expandafter\ifx\csname natexlab\endcsname\relax\def\natexlab#1{#1}\fi
\expandafter\ifx\csname bibnamefont\endcsname\relax
  \def\bibnamefont#1{#1}\fi
\expandafter\ifx\csname bibfnamefont\endcsname\relax
  \def\bibfnamefont#1{#1}\fi
\expandafter\ifx\csname citenamefont\endcsname\relax
  \def\citenamefont#1{#1}\fi
\expandafter\ifx\csname url\endcsname\relax
  \def\url#1{\texttt{#1}}\fi
\expandafter\ifx\csname urlprefix\endcsname\relax\def\urlprefix{URL }\fi
\providecommand{\bibinfo}[2]{#2}
\providecommand{\eprint}[2][]{\url{#2}}

\bibitem[{\citenamefont{Chandrasekhar}(1970{\natexlab{a}})}]{Chandra1}
\bibinfo{author}{\bibfnamefont{S.}~\bibnamefont{Chandrasekhar}},
  \bibinfo{journal}{Phys. Rev. Lett.} \textbf{\bibinfo{volume}{24}},
  \bibinfo{pages}{611} (\bibinfo{year}{1970}{\natexlab{a}}).

\bibitem[{\citenamefont{Chandrasekhar}(1970{\natexlab{b}})}]{Chandra2}
\bibinfo{author}{\bibfnamefont{S.}~\bibnamefont{Chandrasekhar}},
  \bibinfo{journal}{Astrophys. J.} \textbf{\bibinfo{volume}{161}},
  \bibinfo{pages}{561} (\bibinfo{year}{1970}{\natexlab{b}}).

\bibitem[{\citenamefont{Friedman and Schutz}(1978)}]{Frid1}
\bibinfo{author}{\bibfnamefont{J.}~\bibnamefont{Friedman}} \bibnamefont{and}
  \bibinfo{author}{\bibfnamefont{B.}~\bibnamefont{Schutz}},
  \bibinfo{journal}{Astrophys. J.} \textbf{\bibinfo{volume}{222}},
  \bibinfo{pages}{281} (\bibinfo{year}{1978}).

\bibitem[{\citenamefont{Andersson}(1998)}]{Andersson}
\bibinfo{author}{\bibfnamefont{N.}~\bibnamefont{Andersson}},
  \bibinfo{journal}{Astrophys. J.} \textbf{\bibinfo{volume}{502}},
  \bibinfo{pages}{708} (\bibinfo{year}{1998}), \eprint{gr-qc/9706075}.

\bibitem[{\citenamefont{Friedman and Morsink}(1998)}]{Fried2}
\bibinfo{author}{\bibfnamefont{J.~L.} \bibnamefont{Friedman}} \bibnamefont{and}
  \bibinfo{author}{\bibfnamefont{S.~M.} \bibnamefont{Morsink}},
  \bibinfo{journal}{Astrophys. J.} \textbf{\bibinfo{volume}{502}},
  \bibinfo{pages}{714} (\bibinfo{year}{1998}), \eprint{gr-qc/9706073}.

\bibitem[{\citenamefont{Andersson and Kokkotas}(2001)}]{Anderssonreview}
\bibinfo{author}{\bibfnamefont{N.}~\bibnamefont{Andersson}} \bibnamefont{and}
  \bibinfo{author}{\bibfnamefont{K.~D.} \bibnamefont{Kokkotas}},
  \bibinfo{journal}{Int. J. Mod. Phys.} \textbf{\bibinfo{volume}{D10}},
  \bibinfo{pages}{381} (\bibinfo{year}{2001}), \eprint{gr-qc/0010102}.

\bibitem[{\citenamefont{Lindblom}(2000)}]{Lind-lect}
\bibinfo{author}{\bibfnamefont{L.}~\bibnamefont{Lindblom}}
  (\bibinfo{year}{2000}), \eprint{astro-ph/0101136}.

\bibitem[{\citenamefont{Lindblom et~al.}(1998)\citenamefont{Lindblom, Owen, and
  Morsink}}]{Lind1}
\bibinfo{author}{\bibfnamefont{L.}~\bibnamefont{Lindblom}},
  \bibinfo{author}{\bibfnamefont{B.~J.} \bibnamefont{Owen}}, \bibnamefont{and}
  \bibinfo{author}{\bibfnamefont{S.~M.} \bibnamefont{Morsink}},
  \bibinfo{journal}{Phys. Rev. Lett.} \textbf{\bibinfo{volume}{80}},
  \bibinfo{pages}{4843} (\bibinfo{year}{1998}), \eprint{gr-qc/9803053}.

\bibitem[{\citenamefont{Flowers and Itoh}(1976)}]{FlowersItoh1}
\bibinfo{author}{\bibfnamefont{E.}~\bibnamefont{Flowers}} \bibnamefont{and}
  \bibinfo{author}{\bibfnamefont{N.}~\bibnamefont{Itoh}},
  \bibinfo{journal}{Astrophys. J.} \textbf{\bibinfo{volume}{206}},
  \bibinfo{pages}{218} (\bibinfo{year}{1976}).

\bibitem[{\citenamefont{Flowers and Itoh}(1979)}]{FlowersItoh2}
\bibinfo{author}{\bibfnamefont{E.}~\bibnamefont{Flowers}} \bibnamefont{and}
  \bibinfo{author}{\bibfnamefont{N.}~\bibnamefont{Itoh}},
  \bibinfo{journal}{Astrophys. J.} \textbf{\bibinfo{volume}{230}},
  \bibinfo{pages}{847} (\bibinfo{year}{1979}).

\bibitem[{\citenamefont{Sawyer}(1989{\natexlab{a}})}]{Sawyer}
\bibinfo{author}{\bibfnamefont{R.~F.} \bibnamefont{Sawyer}},
  \bibinfo{journal}{Phys. Rev.} \textbf{\bibinfo{volume}{D39}},
  \bibinfo{pages}{3804} (\bibinfo{year}{1989}{\natexlab{a}}).

\bibitem[{\citenamefont{Jones}(2001)}]{Jones1}
\bibinfo{author}{\bibfnamefont{P.~B.} \bibnamefont{Jones}},
  \bibinfo{journal}{Phys. Rev.} \textbf{\bibinfo{volume}{D64}},
  \bibinfo{pages}{084003} (\bibinfo{year}{2001}).

\bibitem[{\citenamefont{Lindblom and Owen}(2002{\natexlab{a}})}]{Lindblom1}
\bibinfo{author}{\bibfnamefont{L.}~\bibnamefont{Lindblom}} \bibnamefont{and}
  \bibinfo{author}{\bibfnamefont{B.~J.} \bibnamefont{Owen}},
  \bibinfo{journal}{Phys. Rev.} \textbf{\bibinfo{volume}{D65}},
  \bibinfo{pages}{063006} (\bibinfo{year}{2002}{\natexlab{a}}),
  \eprint{astro-ph/0110558}.

\bibitem[{\citenamefont{Lindblom and Owen}(2002{\natexlab{b}})}]{Lindblom2}
\bibinfo{author}{\bibfnamefont{L.}~\bibnamefont{Lindblom}} \bibnamefont{and}
  \bibinfo{author}{\bibfnamefont{B.~J.} \bibnamefont{Owen}},
  \bibinfo{journal}{Phys. Rev.} \textbf{\bibinfo{volume}{D65}},
  \bibinfo{pages}{063006} (\bibinfo{year}{2002}{\natexlab{b}}),
  \eprint{astro-ph/0110558}.

\bibitem[{\citenamefont{Drago et~al.}(2005)\citenamefont{Drago, Lavagno, and
  Pagliara}}]{Drago1}
\bibinfo{author}{\bibfnamefont{A.}~\bibnamefont{Drago}},
  \bibinfo{author}{\bibfnamefont{A.}~\bibnamefont{Lavagno}}, \bibnamefont{and}
  \bibinfo{author}{\bibfnamefont{G.}~\bibnamefont{Pagliara}},
  \bibinfo{journal}{Phys. Rev.} \textbf{\bibinfo{volume}{D71}},
  \bibinfo{pages}{103004} (\bibinfo{year}{2005}), \eprint{astro-ph/0312009}.

\bibitem[{\citenamefont{Haensel et~al.}(2000)\citenamefont{Haensel, Levenfish,
  and Yakovlev}}]{Haensel1}
\bibinfo{author}{\bibfnamefont{P.}~\bibnamefont{Haensel}},
  \bibinfo{author}{\bibfnamefont{K.~P.} \bibnamefont{Levenfish}},
  \bibnamefont{and} \bibinfo{author}{\bibfnamefont{D.~G.}
  \bibnamefont{Yakovlev}}, \bibinfo{journal}{Astron. Astrophys.}
  \textbf{\bibinfo{volume}{357}}, \bibinfo{pages}{1157} (\bibinfo{year}{2000}),
  \eprint{astro-ph/0004183}.

\bibitem[{\citenamefont{Haensel et~al.}(2001)\citenamefont{Haensel, Levenfish,
  and Yakovlev}}]{Haensel2}
\bibinfo{author}{\bibfnamefont{P.}~\bibnamefont{Haensel}},
  \bibinfo{author}{\bibfnamefont{K.~P.} \bibnamefont{Levenfish}},
  \bibnamefont{and} \bibinfo{author}{\bibfnamefont{D.~G.}
  \bibnamefont{Yakovlev}}, \bibinfo{journal}{Astron. Astrophys.}
  \textbf{\bibinfo{volume}{372}}, \bibinfo{pages}{130} (\bibinfo{year}{2001}),
  \eprint{astro-ph/0103290}.

\bibitem[{\citenamefont{Chatterjee and Bandyopadhyay}(2006)}]{Chat}
\bibinfo{author}{\bibfnamefont{D.}~\bibnamefont{Chatterjee}} \bibnamefont{and}
  \bibinfo{author}{\bibfnamefont{D.}~\bibnamefont{Bandyopadhyay}},
  \bibinfo{journal}{Phys. Rev.} \textbf{\bibinfo{volume}{D74}},
  \bibinfo{pages}{023003} (\bibinfo{year}{2006}), \eprint{astro-ph/0602538}.

\bibitem[{\citenamefont{Sawyer}(1989{\natexlab{b}})}]{Sawyer2}
\bibinfo{author}{\bibfnamefont{R.~F.} \bibnamefont{Sawyer}},
  \bibinfo{journal}{Phys. Lett.} \textbf{\bibinfo{volume}{B233}},
  \bibinfo{pages}{412} (\bibinfo{year}{1989}{\natexlab{b}}).

\bibitem[{\citenamefont{Madsen}(1992)}]{Madsen}
\bibinfo{author}{\bibfnamefont{J.}~\bibnamefont{Madsen}},
  \bibinfo{journal}{Phys. Rev.} \textbf{\bibinfo{volume}{D46}},
  \bibinfo{pages}{3290} (\bibinfo{year}{1992}).

\bibitem[{\citenamefont{Ivanenko and Kurdgelaidze}(1965)}]{Ivanenko1965}
\bibinfo{author}{\bibfnamefont{D.}~\bibnamefont{Ivanenko}} \bibnamefont{and}
  \bibinfo{author}{\bibfnamefont{D.~F.} \bibnamefont{Kurdgelaidze}},
  \bibinfo{journal}{Astrofiz.} \textbf{\bibinfo{volume}{1}},
  \bibinfo{pages}{479} (\bibinfo{year}{1965}).

\bibitem[{\citenamefont{Ivanenko and Kurdgelaidze}(1969)}]{Ivanenko1969}
\bibinfo{author}{\bibfnamefont{D.}~\bibnamefont{Ivanenko}} \bibnamefont{and}
  \bibinfo{author}{\bibfnamefont{D.~F.} \bibnamefont{Kurdgelaidze}},
  \bibinfo{journal}{Lett. Nuovo Cim.} \textbf{\bibinfo{volume}{2}},
  \bibinfo{pages}{13} (\bibinfo{year}{1969}).

\bibitem[{\citenamefont{Itoh}(1970)}]{Itoh1970}
\bibinfo{author}{\bibfnamefont{N.}~\bibnamefont{Itoh}}, \bibinfo{journal}{Prog.
  Theor. Phys.} \textbf{\bibinfo{volume}{44}}, \bibinfo{pages}{291}
  (\bibinfo{year}{1970}).

\bibitem[{\citenamefont{Iachello et~al.}(1974)\citenamefont{Iachello, Langer,
  and Lande}}]{Iachello1974}
\bibinfo{author}{\bibfnamefont{F.}~\bibnamefont{Iachello}},
  \bibinfo{author}{\bibfnamefont{W.~D.} \bibnamefont{Langer}},
  \bibnamefont{and} \bibinfo{author}{\bibfnamefont{A.}~\bibnamefont{Lande}},
  \bibinfo{journal}{Nucl. Phys. A} \textbf{\bibinfo{volume}{219}},
  \bibinfo{pages}{612} (\bibinfo{year}{1974}).

\bibitem[{\citenamefont{Collins and Perry}(1975)}]{Collins1975}
\bibinfo{author}{\bibfnamefont{J.~C.} \bibnamefont{Collins}} \bibnamefont{and}
  \bibinfo{author}{\bibfnamefont{M.~J.} \bibnamefont{Perry}},
  \bibinfo{journal}{Phys. Rev. Lett.} \textbf{\bibinfo{volume}{34}},
  \bibinfo{pages}{1353} (\bibinfo{year}{1975}).

\bibitem[{\citenamefont{Rajagopal and Wilczek}(2000)}]{RajWil}
\bibinfo{author}{\bibfnamefont{K.}~\bibnamefont{Rajagopal}} \bibnamefont{and}
  \bibinfo{author}{\bibfnamefont{F.}~\bibnamefont{Wilczek}}
  (\bibinfo{year}{2000}), \eprint{hep-ph/0011333}.

\bibitem[{\citenamefont{Alford}(2001)}]{Alfordreview}
\bibinfo{author}{\bibfnamefont{M.~G.} \bibnamefont{Alford}},
  \bibinfo{journal}{Ann. Rev. Nucl. Part. Sci.} \textbf{\bibinfo{volume}{51}},
  \bibinfo{pages}{131} (\bibinfo{year}{2001}), \eprint{hep-ph/0102047}.

\bibitem[{\citenamefont{Reddy}(2002)}]{Reddy2002}
\bibinfo{author}{\bibfnamefont{S.}~\bibnamefont{Reddy}}, \bibinfo{journal}{Acta
  Phys. Polon.} \textbf{\bibinfo{volume}{B33}}, \bibinfo{pages}{4101}
  (\bibinfo{year}{2002}), \eprint{nucl-th/0211045}.

\bibitem[{\citenamefont{Rischke}(2004)}]{Rischke2003}
\bibinfo{author}{\bibfnamefont{D.~H.} \bibnamefont{Rischke}},
  \bibinfo{journal}{Prog. Part. Nucl. Phys.} \textbf{\bibinfo{volume}{52}},
  \bibinfo{pages}{197} (\bibinfo{year}{2004}), \eprint{nucl-th/0305030}.

\bibitem[{\citenamefont{Buballa}(2005)}]{Buballa2003}
\bibinfo{author}{\bibfnamefont{M.}~\bibnamefont{Buballa}},
  \bibinfo{journal}{Phys. Rept.} \textbf{\bibinfo{volume}{407}},
  \bibinfo{pages}{205} (\bibinfo{year}{2005}), \eprint{hep-ph/0402234}.

\bibitem[{\citenamefont{Huang}(2006)}]{Huangreview}
\bibinfo{author}{\bibfnamefont{M.}~\bibnamefont{Huang}}, \bibinfo{journal}{Int.
  J. Mod. Phys.} \textbf{\bibinfo{volume}{A21}}, \bibinfo{pages}{910}
  (\bibinfo{year}{2006}), \eprint{hep-ph/0509177}.

\bibitem[{\citenamefont{Shovkovy}(2005)}]{Shovkovy2004}
\bibinfo{author}{\bibfnamefont{I.~A.} \bibnamefont{Shovkovy}},
  \bibinfo{journal}{Found. Phys.} \textbf{\bibinfo{volume}{35}},
  \bibinfo{pages}{1309} (\bibinfo{year}{2005}), \eprint{nucl-th/0410091}.

\bibitem[{\citenamefont{Shovkovy and Ellis}(2002)}]{ShovEllis1}
\bibinfo{author}{\bibfnamefont{I.~A.} \bibnamefont{Shovkovy}} \bibnamefont{and}
  \bibinfo{author}{\bibfnamefont{P.~J.} \bibnamefont{Ellis}},
  \bibinfo{journal}{Phys. Rev.} \textbf{\bibinfo{volume}{C66}},
  \bibinfo{pages}{015802} (\bibinfo{year}{2002}), \eprint{hep-ph/0204132}.

\bibitem[{\citenamefont{Shovkovy and Ellis}(2003)}]{ShovEllis2}
\bibinfo{author}{\bibfnamefont{I.~A.} \bibnamefont{Shovkovy}} \bibnamefont{and}
  \bibinfo{author}{\bibfnamefont{P.~J.} \bibnamefont{Ellis}},
  \bibinfo{journal}{Phys. Rev.} \textbf{\bibinfo{volume}{C67}},
  \bibinfo{pages}{048801} (\bibinfo{year}{2003}), \eprint{hep-ph/0211049}.

\bibitem[{\citenamefont{Madsen}(2000)}]{Madsenprl2}
\bibinfo{author}{\bibfnamefont{J.}~\bibnamefont{Madsen}},
  \bibinfo{journal}{Phys. Rev. Lett.} \textbf{\bibinfo{volume}{85}},
  \bibinfo{pages}{10} (\bibinfo{year}{2000}), \eprint{astro-ph/9912418}.

\bibitem[{\citenamefont{Manuel et~al.}(2005)\citenamefont{Manuel, Dobado, and
  Llanes-Estrada}}]{Manuel}
\bibinfo{author}{\bibfnamefont{C.}~\bibnamefont{Manuel}},
  \bibinfo{author}{\bibfnamefont{A.}~\bibnamefont{Dobado}}, \bibnamefont{and}
  \bibinfo{author}{\bibfnamefont{F.~J.} \bibnamefont{Llanes-Estrada}},
  \bibinfo{journal}{JHEP} \textbf{\bibinfo{volume}{09}}, \bibinfo{pages}{076}
  (\bibinfo{year}{2005}), \eprint{hep-ph/0406058}.

\bibitem[{\citenamefont{Alford and Schmitt}(2007)}]{Alford:2006gy}
\bibinfo{author}{\bibfnamefont{M.~G.} \bibnamefont{Alford}} \bibnamefont{and}
  \bibinfo{author}{\bibfnamefont{A.}~\bibnamefont{Schmitt}},
  \bibinfo{journal}{J. Phys. G} \textbf{\bibinfo{volume}{34}},
  \bibinfo{pages}{67} (\bibinfo{year}{2007}), \eprint{nucl-th/0608019}.

\bibitem[{\citenamefont{Iwasaki and Iwado}(1995)}]{Iwasaki}
\bibinfo{author}{\bibfnamefont{M.}~\bibnamefont{Iwasaki}} \bibnamefont{and}
  \bibinfo{author}{\bibfnamefont{T.}~\bibnamefont{Iwado}},
  \bibinfo{journal}{Phys. Lett.} \textbf{\bibinfo{volume}{B350}},
  \bibinfo{pages}{163} (\bibinfo{year}{1995}).

\bibitem[{\citenamefont{Pisarski and Rischke}(2000)}]{PD1}
\bibinfo{author}{\bibfnamefont{R.~D.} \bibnamefont{Pisarski}} \bibnamefont{and}
  \bibinfo{author}{\bibfnamefont{D.~H.} \bibnamefont{Rischke}},
  \bibinfo{journal}{Phys. Rev.} \textbf{\bibinfo{volume}{D61}},
  \bibinfo{pages}{074017} (\bibinfo{year}{2000}), \eprint{nucl-th/9910056}.

\bibitem[{\citenamefont{Sch{\"a}fer}(2000)}]{SchaferSpin1}
\bibinfo{author}{\bibfnamefont{T.}~\bibnamefont{Sch{\"a}fer}},
  \bibinfo{journal}{Phys. Rev.} \textbf{\bibinfo{volume}{D62}},
  \bibinfo{pages}{094007} (\bibinfo{year}{2000}), \eprint{hep-ph/0006034}.

\bibitem[{\citenamefont{Schmitt et~al.}(2002)\citenamefont{Schmitt, Wang, and
  Rischke}}]{Schmitt:2002sc}
\bibinfo{author}{\bibfnamefont{A.}~\bibnamefont{Schmitt}},
  \bibinfo{author}{\bibfnamefont{Q.}~\bibnamefont{Wang}}, \bibnamefont{and}
  \bibinfo{author}{\bibfnamefont{D.~H.} \bibnamefont{Rischke}},
  \bibinfo{journal}{Phys. Rev.} \textbf{\bibinfo{volume}{D66}},
  \bibinfo{pages}{114010} (\bibinfo{year}{2002}), \eprint{nucl-th/0209050}.

\bibitem[{\citenamefont{Buballa et~al.}(2003)\citenamefont{Buballa, Hosek, and
  Oertel}}]{BuballaSpin1}
\bibinfo{author}{\bibfnamefont{M.}~\bibnamefont{Buballa}},
  \bibinfo{author}{\bibfnamefont{J.}~\bibnamefont{Hosek}}, \bibnamefont{and}
  \bibinfo{author}{\bibfnamefont{M.}~\bibnamefont{Oertel}},
  \bibinfo{journal}{Phys. Rev. Lett.} \textbf{\bibinfo{volume}{90}},
  \bibinfo{pages}{182002} (\bibinfo{year}{2003}), \eprint{hep-ph/0204275}.

\bibitem[{\citenamefont{Alford et~al.}(2003)\citenamefont{Alford, Bowers,
  Cheyne, and Cowan}}]{AlfordSpin1}
\bibinfo{author}{\bibfnamefont{M.~G.} \bibnamefont{Alford}},
  \bibinfo{author}{\bibfnamefont{J.~A.} \bibnamefont{Bowers}},
  \bibinfo{author}{\bibfnamefont{J.~M.} \bibnamefont{Cheyne}},
  \bibnamefont{and} \bibinfo{author}{\bibfnamefont{G.~A.} \bibnamefont{Cowan}},
  \bibinfo{journal}{Phys. Rev.} \textbf{\bibinfo{volume}{D67}},
  \bibinfo{pages}{054018} (\bibinfo{year}{2003}), \eprint{hep-ph/0210106}.

\bibitem[{\citenamefont{Schmitt}(2005)}]{SchmittSpin1}
\bibinfo{author}{\bibfnamefont{A.}~\bibnamefont{Schmitt}},
  \bibinfo{journal}{Phys. Rev.} \textbf{\bibinfo{volume}{D71}},
  \bibinfo{pages}{054016} (\bibinfo{year}{2005}), \eprint{nucl-th/0412033}.

\bibitem[{\citenamefont{Grigorian et~al.}(2005)\citenamefont{Grigorian,
  Blaschke, and Voskresensky}}]{Grigorian:2004jq}
\bibinfo{author}{\bibfnamefont{H.}~\bibnamefont{Grigorian}},
  \bibinfo{author}{\bibfnamefont{D.}~\bibnamefont{Blaschke}}, \bibnamefont{and}
  \bibinfo{author}{\bibfnamefont{D.}~\bibnamefont{Voskresensky}},
  \bibinfo{journal}{Phys. Rev.} \textbf{\bibinfo{volume}{C71}},
  \bibinfo{pages}{045801} (\bibinfo{year}{2005}), \eprint{astro-ph/0411619}.

\bibitem[{\citenamefont{Sa'd et~al.}(2006)\citenamefont{Sa'd, Shovkovy, and
  Rischke}}]{SadShR}
\bibinfo{author}{\bibfnamefont{B.~A.} \bibnamefont{Sa'd}},
  \bibinfo{author}{\bibfnamefont{I.~A.} \bibnamefont{Shovkovy}},
  \bibnamefont{and} \bibinfo{author}{\bibfnamefont{D.~H.}
  \bibnamefont{Rischke}}, \bibinfo{journal}{in preparation}
  (\bibinfo{year}{2006}).

\bibitem[{\citenamefont{Schmitt et~al.}(2006)\citenamefont{Schmitt, Shovkovy,
  and Wang}}]{SSW2}
\bibinfo{author}{\bibfnamefont{A.}~\bibnamefont{Schmitt}},
  \bibinfo{author}{\bibfnamefont{I.~A.} \bibnamefont{Shovkovy}},
  \bibnamefont{and} \bibinfo{author}{\bibfnamefont{Q.}~\bibnamefont{Wang}},
  \bibinfo{journal}{Phys. Rev.} \textbf{\bibinfo{volume}{D73}},
  \bibinfo{pages}{034012} (\bibinfo{year}{2006}), \eprint{hep-ph/0510347}.

\bibitem[{\citenamefont{Wang et~al.}(2006)\citenamefont{Wang, Wang, and
  Wu}}]{Wang2006}
\bibinfo{author}{\bibfnamefont{Q.}~\bibnamefont{Wang}},
  \bibinfo{author}{\bibfnamefont{Z.-g.} \bibnamefont{Wang}}, \bibnamefont{and}
  \bibinfo{author}{\bibfnamefont{J.}~\bibnamefont{Wu}}, \bibinfo{journal}{Phys.
  Rev.} \textbf{\bibinfo{volume}{D74}}, \bibinfo{pages}{014021}
  (\bibinfo{year}{2006}), \eprint{hep-ph/0605092}.

\bibitem[{\citenamefont{Bhattacharyya et~al.}(1977)\citenamefont{Bhattacharyya,
  Pethick, and Smith}}]{prb15Bhatta}
\bibinfo{author}{\bibfnamefont{P.}~\bibnamefont{Bhattacharyya}},
  \bibinfo{author}{\bibfnamefont{C.~J.} \bibnamefont{Pethick}},
  \bibnamefont{and} \bibinfo{author}{\bibfnamefont{H.}~\bibnamefont{Smith}},
  \bibinfo{journal}{Phys. Rev. B} \textbf{\bibinfo{volume}{15}},
  \bibinfo{pages}{3367} (\bibinfo{year}{1977}).

\bibitem[{\citenamefont{Landau and Lifshitz}(1987)}]{LL6}
\bibinfo{author}{\bibfnamefont{L.~D.} \bibnamefont{Landau}} \bibnamefont{and}
  \bibinfo{author}{\bibfnamefont{E.~M.} \bibnamefont{Lifshitz}},
  \emph{\bibinfo{title}{Fluid mechanics}} (\bibinfo{publisher}{Pergamon Press},
  \bibinfo{year}{1987}).

\bibitem[{\citenamefont{Son}(2007)}]{Son:2005tj}
\bibinfo{author}{\bibfnamefont{D.~T.} \bibnamefont{Son}},
  \bibinfo{journal}{Phys. Rev. Lett.} \textbf{\bibinfo{volume}{98}},
  \bibinfo{pages}{020604} (\bibinfo{year}{2007}), \eprint{cond-mat/0511721}.

\bibitem[{\citenamefont{Graham}(1974)}]{Graham1974}
\bibinfo{author}{\bibfnamefont{R.}~\bibnamefont{Graham}},
  \bibinfo{journal}{Phys. Rev. Lett.} \textbf{\bibinfo{volume}{33}},
  \bibinfo{pages}{1431} (\bibinfo{year}{1974}).

\bibitem[{\citenamefont{Freedman and McLerran}(1977)}]{Freedman1976}
\bibinfo{author}{\bibfnamefont{B.~A.} \bibnamefont{Freedman}} \bibnamefont{and}
  \bibinfo{author}{\bibfnamefont{L.~D.} \bibnamefont{McLerran}},
  \bibinfo{journal}{Phys. Rev.} \textbf{\bibinfo{volume}{D16}},
  \bibinfo{pages}{1169} (\bibinfo{year}{1977}).

\bibitem[{\citenamefont{Baluni}(1978)}]{Baluni1977}
\bibinfo{author}{\bibfnamefont{V.}~\bibnamefont{Baluni}},
  \bibinfo{journal}{Phys. Rev.} \textbf{\bibinfo{volume}{D17}},
  \bibinfo{pages}{2092} (\bibinfo{year}{1978}).

\bibitem[{\citenamefont{Fraga and Romatschke}(2005)}]{Fraga2004}
\bibinfo{author}{\bibfnamefont{E.~S.} \bibnamefont{Fraga}} \bibnamefont{and}
  \bibinfo{author}{\bibfnamefont{P.}~\bibnamefont{Romatschke}},
  \bibinfo{journal}{Phys. Rev.} \textbf{\bibinfo{volume}{D71}},
  \bibinfo{pages}{105014} (\bibinfo{year}{2005}), \eprint{hep-ph/0412298}.

\bibitem[{\citenamefont{Sch{\"a}fer}(2006)}]{SchaferChapter}
\bibinfo{author}{\bibfnamefont{T.}~\bibnamefont{Sch{\"a}fer}}
  (\bibinfo{year}{2006}), \eprint{nucl-th/0602067}.

\bibitem[{\citenamefont{Iwamoto}(1980)}]{Iwamoto}
\bibinfo{author}{\bibfnamefont{N.}~\bibnamefont{Iwamoto}},
  \bibinfo{journal}{Phys. Rev. Lett.} \textbf{\bibinfo{volume}{44}},
  \bibinfo{pages}{1637} (\bibinfo{year}{1980}).

\bibitem[{\citenamefont{Aguilera et~al.}(2005)\citenamefont{Aguilera, Blaschke,
  Buballa, and Yudichev}}]{ABBY}
\bibinfo{author}{\bibfnamefont{D.~N.} \bibnamefont{Aguilera}},
  \bibinfo{author}{\bibfnamefont{D.}~\bibnamefont{Blaschke}},
  \bibinfo{author}{\bibfnamefont{M.}~\bibnamefont{Buballa}}, \bibnamefont{and}
  \bibinfo{author}{\bibfnamefont{V.~L.} \bibnamefont{Yudichev}},
  \bibinfo{journal}{Phys. Rev.} \textbf{\bibinfo{volume}{D72}},
  \bibinfo{pages}{034008} (\bibinfo{year}{2005}), \eprint{hep-ph/0503288}.

\bibitem[{\citenamefont{Buballa}(2006)}]{BuballaComm}
\bibinfo{author}{\bibfnamefont{M.}~\bibnamefont{Buballa}},
  \bibinfo{journal}{private communication}  (\bibinfo{year}{2006}).

\end{thebibliography}
\end{document}